\documentclass[runningheads,a4paper]{llncs}

\usepackage{amssymb}
\usepackage{graphicx}

\usepackage{amsmath}
\usepackage{stfloats}
\usepackage{amsfonts}

\usepackage{algorithm,algorithmic}

\usepackage{url}
\urldef{\mailsa}\path|{alfred.hofmann, ursula.barth, ingrid.haas, frank.holzwarth,|
\urldef{\mailsb}\path|anna.kramer, leonie.kunz, christine.reiss, nicole.sator,|
\urldef{\mailsc}\path|erika.siebert-cole, peter.strasser, lncs}@springer.com|
\newcommand{\keywords}[1]{\par\addvspace\baselineskip
\noindent\keywordname\enspace\ignorespaces#1}

\begin{document}

\mainmatter  

\title{Game-theoretic Analysis to Content-adaptive Reversible Watermarking}

\titlerunning{Game-theoretic Analysis to Content-adaptive Reversible Watermarking}

%
%
\author{Hanzhou Wu$^*$ and Xinpeng Zhang}
\authorrunning{Hanzhou Wu and Xinpeng Zhang}

\institute{Shanghai University, Shanghai 200444, China\\
$^*$\email{h.wu.phd@ieee.org}, \email{xzhang@shu.edu.cn}\\
\textbf{manuscript draft - 1st version}
}
%
%

\maketitle

\begin{abstract}
While many games were designed for steganography and robust watermarking, few focused on reversible watermarking. We present a two-encoder game related to the rate-distortion optimization of content-adaptive reversible watermarking. In the game, Alice first hides a payload into a cover. Then, Bob hides another payload into the modified cover. The embedding strategy of Alice affects the embedding capacity of Bob. The embedding strategy of Bob may produce data-extraction errors to Alice. Both want to embed as many pure secret bits as possible, subjected to an upper-bounded distortion. We investigate non-cooperative game and cooperative game between Alice and Bob. When they cooperate with each other, one may consider them as a whole, i.e., an encoder uses a cover for data embedding with two times. When they do not cooperate with each other, the game corresponds to a separable system, i.e., both want to independently hide a payload within the cover, but recovering the cover may need cooperation. We find equilibrium strategies for both players under constraints.
\end{abstract}

\section{Introduction}
Reversible watermarking (RW), also called reversible data hiding (RDH) \cite{tian:de}, \cite{ni:hs}, enables us to embed a secret payload in a cover. Both the embedded payload and the original cover content can be perfectly retrieved from the marked content. RW is helpful in sensitive scenarios requiring no permanent distortion of the cover such as military communication and multimedia archive management. RW is fragile, i.e., one will find it is not authentic if the marked data was tampered. RW can be evaluated by the rate-distortion performance. That is, we always expect to embed as many message bits as possible for a fixed distortion level. In other words, we hope to reduce the distortion as much as possible for a payload.

A straightforward idea to achieve reversibility is applying lossless compression \cite{fridrich:inv}, i.e., an encoder selects the noise-like component of the cover such as the least significant bits (LSBs) of a grayscale image, and then losslessly compress them to reserve additional space. The secret message will be embedded into the reserved space. Though altering the noise-like component does not introduce noticeable visual artifacts, the pure embedding capacity is limited due to the low compression rate. To this end, more efficient algorithms such as difference expansion \cite{tian:de}, histogram shifting \cite{ni:hs} and their variants \cite{sachnev:sp}, \cite{hzwu:PPE}, \cite{hzwu:DCSPF}, are proposed to enlarge capacity or reduce distortion. We will not review these works since it is not the main interest of this paper.

Though current advanced RW algorithms differ from lossless compression intuitively, they are essentially similar to each other. They all aim to find the compressible component of the cover to carry the secret message. The difference is, lossless compression considers cover elements as compressible variables, while the others try to fully exploit the correlations between elements, which can be regarded as a kind of ``semantic lossless compression'' \cite{weiming:equal}.

From the theoretical perspective, studying lossless compression based RW is still desirable, though people have developed a number of new practical algorithms. The theoretical results is helpful to guide or inspire us to design new RW systems, which has motivated the authors to revisit lossless compression based RW in this paper. However, different from traditional works, we use game theory to analyze RW. Unlike previous games designed to robust or saying irreversible watermarking games \cite{game1998}, \cite{titGame2003}, \cite{GameWatermarking}, in the proposed game, there are two players named Alice and Bob, both of whom want to embed a different payload into a specific cover object. It is different from previous practical RW systems, where only one encoder is considered. We point that, our work actually can generalize them. The reason is, Alice and Bob could cooperate with each other. In this way, one may consider Alice and Bob as a whole, indicating that, an encoder will use a cover for data embedding with two times, which corresponds to \emph{multi-layer embedding}. When they do not cooperate with each other, the game between Alice and Bob actually corresponds to a \emph{separable} RW system, i.e., both want to independently hide a payload within the cover, but recovering the cover may need cooperation.

The rest are organized as follows. In Section 2, we review prior arts. In Section 3, we introduce the basic setup of our game model. In Section 4, we analyze the non-cooperative game between Alice and Bob, followed by the cooperative game in Section 5. We conclude this work in Section 6.

\section{Prior Arts}
Though steganography should be distinguished from watermarking, they are correlated to each other. The first work combining game theory and steganography was probably proposed by Ettinger \cite{game1998}. In the work, a zero-sum game between the data hider and the attacker was presented, for which the equilibrium is such a strategy profile that none of the two players wants to profit from adjusting two different partial strategies, for a fixed strategy of the opponent. Moulin and O'Sullivan \cite{titGame2003} characterized the watermarking codes as a capacity game between the data hider and the attacker, in which two models, i.e., private game and public game, of watermarking systems were presented. Baruch and Merhav \cite{GameWatermarking} investigated the capacity and error exponents games of private watermarking games, where the attack channel is completely general and unknown to the hider and receiver.

Cohen and Lapidoth \cite{cohen2002} derived a coding capacity formula of the watermarking game for a Gaussian covertext and squared-error distortions. They showed that the capacities of the public and private watermarking are the same, which is in some analogy to Costa's result \cite{costa1983} on channel coding with side information under the Gaussian quadratic regime. Baruch and Merhav \cite{capPWS2004} extended the public game \cite{titGame2003} by dropping the assumptions that the receiver knows the attack channel and that this channel is memoryless or blockwise memoryless.

Ker \cite{kerGame} introduced a threshold game to batch steganography \cite{kerBatch}, in which a steganographer should decide how to distribute a payload into multiple pieces each embedded into a selective cover. The result indicated that, the optimal strategy of the steganographer is likely to be extreme concentration of the payload into as few as covers as possible, or the payload is spread as thinly as possible.

Sch$\ddot{\text{o}}$ttle and B$\ddot{\text{o}}$hme \cite{game2playersPSRB} defined the heterogeneity as a necessary condition for adaptive steganography, and presented a game-theoretic model for the whole steganographic process including cover generation, adaptive embedding, and a detector which anticipates the adaptivity. Though the model exhibited a unique equilibrium in mixed strategies, it investigated a cover model with only two locations. To this end, Sch$\ddot{\text{o}}$ttle and B$\ddot{\text{o}}$hme \cite{Where2hideBits} further extended the model from the very artificial case to covers sized $n$. Their results are constructive in sense that an equilibrium can be efficiently found for any vector of predictability. More games designed for adaptive steganography can be found in
\cite{jessicaGame2014}, \cite{gameAnalysis2013}, \cite{gameTIFS}.

\section{Game Setup}
\subsection{Description of the Cover}
The cover to be embedded is written as \textbf{X} = $(\textbf{x}_1, \textbf{x}_2, ..., \textbf{x}_l)$, where $\textbf{x}_i = \{x_{i,j}\}_{j=1}^{n}$ represents a binary sequence with the distribution $p_{i,1} = \sum_{j=1}^{n}x_{i,j} / n$ and $p_{i,0} = 1 - p_{i,1}$. It is assumed that, for any $i\neq j$, there has no intersection between $\textbf{x}_i$ and $\textbf{x}_j$. We point that, the term `cover' here actually means the data-embedding channel. For example, if we use the LSBs of an image for data embedding, we can separate the LSB plane from the image, and divide it into disjoint bit-vectors to constitute \textbf{X}. On the one hand, such definition has good generalization for lossless compression based reversible watermarking. On the other hand, it allows us to well model the rate-distortion game between two encoders latter.

\subsection{Rate-distortion Game between Two Players}
In RW, to achieve superior rate-distortion performance, it is generally required that, the secret bits should be preferentially embedded in the smooth area of the cover, rather than the complex one. Since we use binary sequences here, the smoothness of a binary sequence is defined as the information entropy. The smaller the entropy, the more the embeddable bits, implying better rate-distortion performance. We define the smoothness of $\textbf{x}_i$ as $H(p_i)$, where $p_i = \text{min}\{p_{i,0}, p_{i,1}\}\leq \frac{1}{2}$, and
\begin{equation*}
H(p_i) = -p_i\cdot\text{log}_2(p_i)-(1-p_i)\cdot\text{log}_2(1-p_i)
\end{equation*}
means the binary entropy. Without the loss of generality, we assume that
\begin{equation*}
H(p_1) \leq H(p_2) \leq ... \leq H(p_l),
\end{equation*}
meaning that, $\textbf{x}_i$ is more suitable for data embedding than $\textbf{x}_j$, for any $i < j$.

Suppose that, there are two players Alice and Bob, both expecting to hide a payload within \textbf{X}. Alice first hides a payload by modifying the bits throughout \textbf{X} under the constraint that the total amount of distortion is no more than $d$. A strategy for Alice is an $l$-tuple of probabilities $(s_1, s_2, ..., s_l)$, where $s_i \in [0, 1], \forall i\in [1, l]$. It indicates that, for each $i\in [1,l]$, Alice chooses $s_i\times 100 \%$ pixels from $\textbf{x}_i$ with a secret key and losslessly compress the bits. These bit-positions will carry the compressed code and the secret payload.

The resulting data \textbf{Y} will be sent to Bob, who embeds another payload into \textbf{Y} by the same means subjected to $d$ as well, resulting in another marked data \textbf{Z}. A strategy for Bob is therefore another $l$-tuple of probabilities $(t_1, t_2, ..., t_l)$. Though both strategies are real vectors, the number of bit-positions is an integer. We use real numbers for better analysis.

With \textbf{Z}, Bob can directly retrieve the hidden data and his compressed code. However, Alice should correct the errors produced by Bob after data extraction. For Alice, the compressed code for $\textbf{x}_i$ requires $ns_iH(p_i)$ bits. Since the channel capacity for a binary symmetric channel with bit error probability $p$ is $1-H(p)$, the size of embeddable payload of $\textbf{x}_i$ in bits for Alice is therefore
\begin{equation}
A_i(s_i, t_i) = ns_i\left(1 - H(\frac{t_i}{2}) - H(p_i)\right).
\end{equation}
The size of embeddable payload of $\textbf{x}_i$ in bits for Bob is
\begin{equation}
B_i(s_i, t_i) = nt_i\left(1 - H(p_i+\frac{s_i}{2}-p_is_i)\right).
\end{equation}

For maximizing the embedding capacity, a strategy profile $(\textbf{s}, \textbf{t})$ = $(s_1,s_2,...,s_l;$ $t_1,t_2,...,t_l)$ for Alice gives a payoff
\begin{equation}
P_\text{A}(\textbf{s}, \textbf{t}) = \sum_{i=1}^{l}A_i(s_i, t_i),
\end{equation}
and the payoff for Bob is
\begin{equation}
P_\text{B}(\textbf{s}, \textbf{t}) = \sum_{i=1}^{l}B_i(s_i, t_i).
\end{equation}

It can be easily proved that
\begin{equation}
\begin{split}
P_\text{A}(\textbf{s}, \textbf{t})&\leq \sum_{i=1}^{l}ns_i\left ( 1 - H(p_i) \right )\\
&\text{and}~P_\text{B}(\textbf{s}, \textbf{t}) \leq \sum_{i=1}^{l}nt_i\left ( 1 - H(p_i) \right ).
\end{split}
\end{equation}

Both want to maximize their own payoff, while subjected to a bounded distortion. Since data embedding in smooth area gives a larger capacity, intuitively, both will absolutely embed their own payload into $\textbf{x}_i$ with a small index (if there has no opponent). Thus, when introducing competition, as limited and precious resource, the cost of data embedding in $\textbf{x}_i$ will be surely higher than that of $\textbf{x}_j$, for any $i < j$ for both players. Let $\rho(\textbf{x}_i)$ denote the cost (distortion) of flipping any individual bit in $\textbf{x}_i$. It can be therefore assumed that,
\begin{equation}
\rho(\textbf{x}_i) \geq \rho(\textbf{x}_{i+1}), \forall i\in [1,l-1],
\end{equation}
i.e.,
\begin{equation}
\rho(\textbf{x}_i) \propto \frac{1}{H(p_i)}, \forall i\in [1,l].
\end{equation}

We limit us to additive distortion. Namely, we have
\begin{equation}
D_\text{A}(\textbf{s}) = \sum_{i=1}^{l}\frac{ns_i}{2}\rho(\textbf{x}_i) \leq d,
\end{equation}
and
\begin{equation}
D_\text{B}(\textbf{t}) = \sum_{i=1}^{l}\frac{nt_i}{2}\rho(\textbf{x}_i) \leq d.
\end{equation}

By using different $\rho$, sophisticated models of the effects of modifying \textbf{X} may be accommodated in this game. It is pointed that, though there has an embedding order for two players, they both know complete information about the game except the data-embedding key held by the opponent.

\begin{figure}[!t]
  \centering
  \includegraphics[width=4.8in]{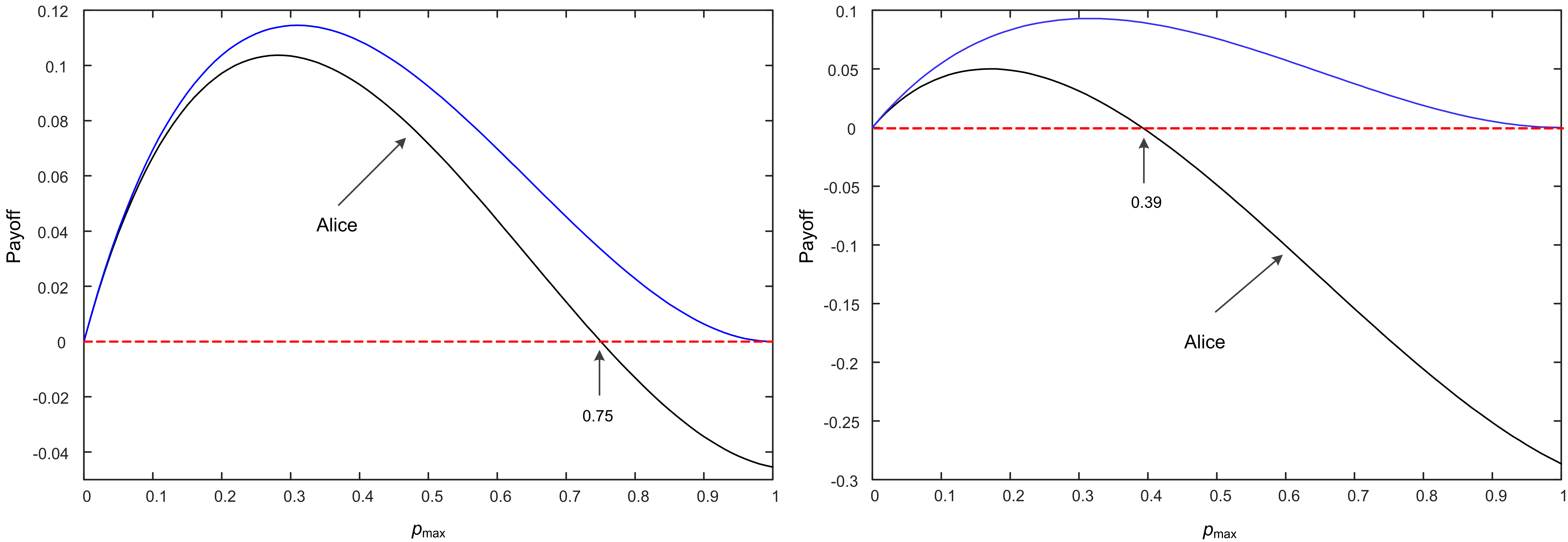}\\
  \caption{Payoffs (divided by $n$) due to different $p_\text{max}$. left: $p_1 = 0.005$, right: $p_1=0.05$.}\label{f1}
\end{figure}

\section{Non-cooperative Game}
When Alice and Bob do not cooperate with each other, an \emph{equilibrium} is a profile $(\textbf{s}^*, \textbf{t}^*)$ = $(s_1^*,s_2^*,...,s_l^*;$ $t_1^*,t_2^*,...,t_l^*)$ that
\begin{equation}
P_\text{A}(\textbf{s}, \textbf{t}^*) \leq P_\text{A}(\textbf{s}^*, \textbf{t}^*),
\end{equation}
and
\begin{equation}
P_\text{B}(\textbf{s}^*, \textbf{t}) \leq P_\text{B}(\textbf{s}^*, \textbf{t}^*).
\end{equation}

\subsection{Case $l = 1$}
We first analyze the case $l = 1$. It can be easily obtained from Eq. (8, 9) that
\begin{equation}
s_1 \leq p_\text{max} = \text{min}\{1, \frac{2d}{n\rho(\textbf{x}_1)}\}~\text{and}~t_1 \leq p_\text{max}.
\end{equation}

It is observed from Eq. (2) that, no matter what strategy Alice takes, Bob will always choose $t_1 = p_\text{max}$ since it achieves the maximum payoff. For a fixed $t_1$, the optimal response for Alice will be $s_1 = p_\text{max}$. Therefore, the equilibrium is uniquely $(\textbf{s}^*, \textbf{t}^*) = (p_\text{max}, p_\text{max})$. Fig. 1 shows the equilibrium payoffs due to different $p_\text{max}$. It is observed that, the payoff of Bob is never less than Alice. Since Alice needs correct errors produced by Bob, when the distortion threshold is larger than a threshold (which allows Bob to produce more errors), Alice may not embed any extra bits (i.e., the payoff is negative), e.g., in Fig. 1, when $p_\text{max} > 0.75$ for the left case, Alice cannot embed a pure payload.

\subsection{Case $l = 2$}
Let $T_i = 1 - H(\frac{t_i}{2}) - H(p_i)$ and $S_i = 1 - H(p_i+\frac{s_i}{2}-p_is_i)$, we have
\begin{equation}
P_\text{A}(\textbf{s}, \textbf{t}) = \sum_{i=1}^{l}ns_iT_i~\text{and}~P_\text{B}(\textbf{s}, \textbf{t}) = \sum_{i=1}^{l}nt_iS_i.
\end{equation}

Neither Alice or Bob wants to deviate from an equilibrium. In case $l = 2$, to find an equilibrium, we must insure that Bob does not profit from readjusting $t_1$ and $t_2$ for a fixed \textbf{s}. Similarly, Alice does not profit from readjusting $s_1$ and $s_2$ for a fixed \textbf{t}. Therefore we have
\begin{equation}
\frac{\partial P_\text{B}(\textbf{s}, \textbf{t})}{\partial t_i}(t_i^*) = 0~\text{and}~\frac{\partial P_\text{A}(\textbf{s}, \textbf{t})}{\partial s_i}(s_i^*) = 0, \forall i\in \{1,2\},
\end{equation}
from which we can obtain
\begin{equation}
\frac{S_1}{S_2} = \frac{T_1}{T_2} = \frac{\rho(\textbf{x}_1)}{\rho(\textbf{x}_2)}.
\end{equation}

When $\frac{\rho(\textbf{x}_1)}{\rho(\textbf{x}_2)}$ is specified at the very beginning, it is straightforward to draw out all strategy profiles meeting Eq. (15). For the `sub-cover' $\textbf{x}_i$, let $d_i^\text{A}$ and $d_i^\text{B}$ be the bounded distortion for both players, i.e.,
\begin{equation}
\frac{ns_i}{2}\rho(\textbf{x}_i) \leq d_i^\text{A}~\text{and}~\frac{nt_i}{2}\rho(\textbf{x}_i) \leq d_i^\text{B}.
\end{equation}

The optimal `sub-strategy' profile is therefore determined as $(\text{min}\{1, \frac{2d_i^\text{A}}{n\rho(\textbf{x}_i)}\}$, $\text{min}\{1, \frac{2d_i^\text{B}}{n\rho(\textbf{x}_i)}\})$. It implies that, the optimal response for an individual player is only dependent of his or her own distortion distribution. And, the distortion upper-bound itself for a sub-cover gives the maximum `sub-payoff'. It can be derived that, $(\textbf{s}^*, \textbf{t}^*)$ can be determined by moving a specified initial strategy profile towards the corresponding direction until the distortion bound is reached.

We take Fig. 2 for example, where $p_1 = 0.005$, $p_2=0.05$, $\frac{\rho(\textbf{x}_1)}{\rho(\textbf{x}_2)} = 2$. The left draws out all strategies that meet Eq. (15). It reduces the raw strategy space for determining the final equilibrium. By sampling a set of points with a small step (in our simulation, we set its value as $0.1\times 10^3$) on both curves, we can produce two strategy sequences, each element in which can be orderly indexed by a value.
Thus, we can draw out all strategy profiles as the right figure shown in Fig. 2. It is observed that, fixing the strategy of either player, the payoff curve of the other player is a strictly monotone increasing function. It means that, the equilibrium will be such a strategy profile that maximizes the distortion, which can be determined during the process of moving from a start point to the end point, as shown in the left. Notice that, we always force Bob to allow Alice to embed a non-negative pure payload in arbitrary sub-cover; otherwise, the game is unfair to Alice. Namely, it is always required that
\begin{equation}
1 - H(t_i^*/2) \geq H(p_i),\forall i\in [1, l].
\end{equation}

\begin{figure}[!t]
  \centering
  \includegraphics[width=4.8in]{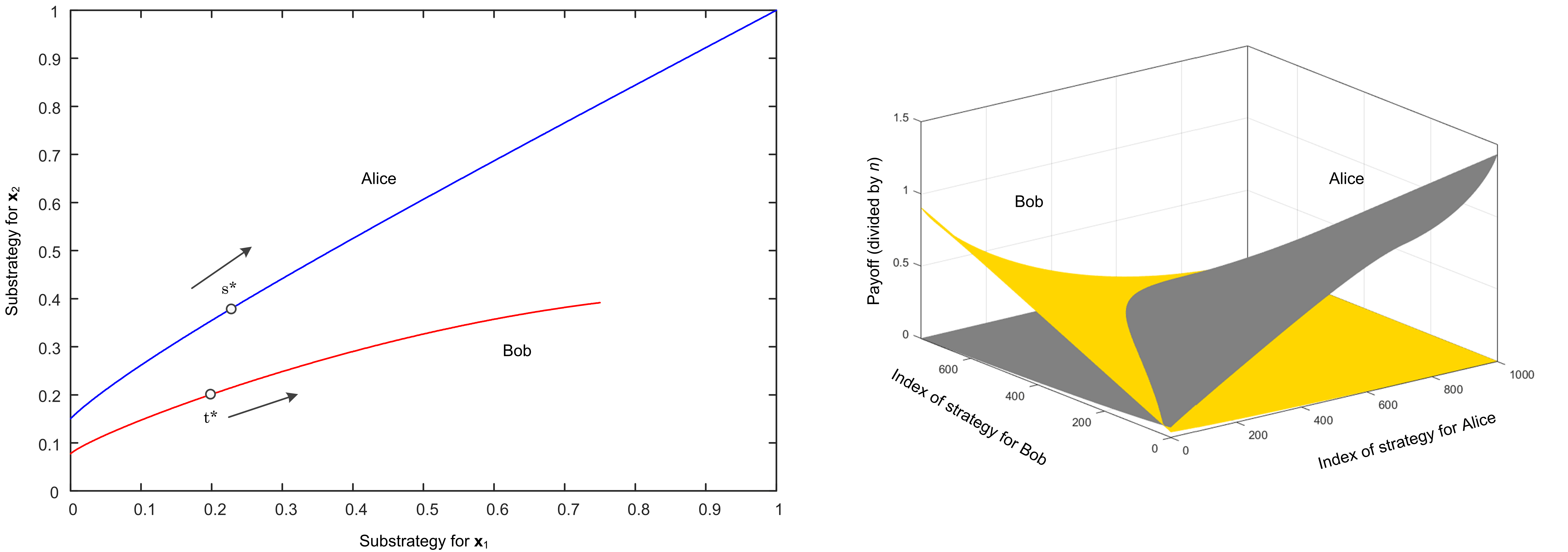}\\
  \caption{An example of equilibrium in case $l=2$: $p_1 = 0.005$, $p_2=0.05$, $\frac{\rho(\textbf{x}_1)}{\rho(\textbf{x}_2)} = 2$.}\label{f2}
\end{figure}

\subsection{Case $l > 2$}
Similarly, in case $l > 2$, neither Alice or Bob deviates from an equilibrium. We fix all components of \textbf{t} except for $t_j$ and $t_k~(j\neq k)$, and distribute the distortion
\begin{equation}
d_\text{B}^{j,k}=d-\sum_{i\neq j,k}\frac{nt_i}{2}\rho(\textbf{x}_i)
\end{equation}
between these two components. Let
\begin{equation}
B_{j,k}(t_j)=nt_jS_j+nt_kS_k=nt_jS_j+\frac{2d_\text{B}^{j,k}-nt_j\rho(\textbf{x}_j)}{\rho(\textbf{x}_k)}S_k.
\end{equation}

To find an equilibrium, we must insure that Bob does not profit from readjusting $t_j$ and $t_k$ for a fixed \textbf{s}. Therefore we have
\begin{equation}
\frac{\partial B_{j,k}}{\partial t_j}(t_j^*)=0
\end{equation}
for an equilibrium $t_j^*$. We continue to fix all the components of \textbf{s} except for $s_j$ and $s_k$ where $1\leq j < k \leq l$, and distribute the remaining distortion
\begin{equation}
d_\text{A}^{j,k}=d-\sum_{i\neq j,k}\frac{ns_i}{2}\rho(\textbf{x}_i)
\end{equation}
between these two components. Let
\begin{equation}
A_{j,k}(s_j)=ns_jT_j+ns_kT_k=ns_jT_j+\frac{2d_\text{A}^{j,k}-ns_j\rho(\textbf{x}_j)}{\rho(\textbf{x}_k)}T_k.
\end{equation}
Since Alice does not profit from readjusting $s_j$ and $s_k$, we have
\begin{equation}
\frac{\partial A_{j,k}}{\partial s_j}(s_j^*)=0
\end{equation}
for an equilibrium $s_j^*$. Thus, according to Eq. (20, 23), we have
\begin{equation}
\frac{S_j}{S_k} = \frac{T_j}{T_k} = \frac{\rho(\textbf{x}_j)}{\rho(\textbf{x}_k)} \geq 1, \forall j < k.
\end{equation}
Suppose that, $S_1 = \alpha\in [0, 1]$ and $T_1 = \beta \in [0, 1]$, we have
\begin{equation}
S_i = \frac{\rho(\textbf{x}_i)}{\rho(\textbf{x}_1)}\alpha~\text{and}~T_i = \frac{\rho(\textbf{x}_i)}{\rho(\textbf{x}_1)}\beta.
\end{equation}

It can be inferred that, the equilibrium $(\textbf{s}^*, \textbf{t}^*)$ should correspond to such a pair $(\alpha^*,\beta^*) \in [0,1]^2$ that,
\begin{equation}
s_i^* = \frac{H^{-1}\left(1-\alpha^*\rho(\textbf{x}_i)/\rho(\textbf{x}_1)\right)-p_i}{1/2-p_i},
\end{equation}
\begin{equation}
t_i^* = 2H^{-1}\left(1-H(p_i)-\beta^*\rho(\textbf{x}_i)/\rho(\textbf{x}_1)\right),
\end{equation}
and
\begin{equation}
\sum_{i=1}^{l}\frac{ns_i^*}{2}\rho(\textbf{x}_i) = \sum_{i=1}^{l}\frac{nt_i^*}{2}\rho(\textbf{x}_i) = d.
\end{equation}
which can be effectively solved by \emph{binary search} since the data-embedding distortion functions for both players are strictly monotone w.r.t. $(\alpha,\beta)$. Notice that, the binary search operation is used twice for determining $\textbf{s}^*$ and $\textbf{t}^*$ respectively.

\section{Cooperative Game}
When Alice and Bob cooperate with each other, the payoff can be rewritten as
\begin{equation}
P(\textbf{s},\textbf{t}) = \text{min} \left\{P_\text{A}(\textbf{s}, \textbf{t}), P_\text{B}(\textbf{s}, \textbf{t})\right\},
\end{equation}
for which an equilibrium is such a pair $(\textbf{s}^*, \textbf{t}^*)$ that \emph{preferentially} meets:
\begin{equation}
\text{max}\left \{P(\textbf{s}^*,\textbf{t}), P(\textbf{s},\textbf{t}^*)\right \} \leq P(\textbf{s}^*, \textbf{t}^*)
\end{equation}
and then meets:
\begin{equation}
\text{max}\left \{P_\text{A}(\textbf{s}^*,\textbf{t}), P_\text{B}(\textbf{s},\textbf{t}^*)\right \} \leq \text{max} \left\{P_\text{A}(\textbf{s}^*, \textbf{t}^*), P_\text{B}(\textbf{s}^*, \textbf{t}^*)\right\}.
\end{equation}

Unlike other cooperative games, to find the equilibrium, we have to consider the side information shared between Alice and Bob. That is, we have to analyze two situations below:

\begin{itemize}
  \item Case 1: They share the key of choosing the bit-positions to be embedded.
  \item Case 2: They do not share the key mentioned above.
\end{itemize}

Case 1 corresponds to the double embedding strategy commonly used in RW. In this case, we can consider Alice and Bob as a whole. When Alice chooses $ns_i$ positions in $\textbf{x}_i$, Bob can use the rest positions. We thus have $s_i+t_i\leq 1,\forall i\in [1,l]$. It indicates that, Bob will not produce extraction errors. Therefore,
\begin{equation}
\begin{split}
P(\textbf{s}, \textbf{t}) &= \sum_{i=1}^{l}n\cdot\text{min}\{s_i, t_i\}\cdot\left(1 - H(p_i)\right)\\
&\leq \sum_{i=1}^{l}n\cdot\frac{s_i+t_i}{2}\cdot\left(1 - H(p_i)\right),
\end{split}
\end{equation}
which allows
\begin{equation}
d\leq d_\text{max} = \sum_{i=1}^{l}\frac{n}{2}\rho(\textbf{x}_i).
\end{equation}

It can be then inferred that, $\textbf{s}^*$ = $\textbf{t}^*$ = ($\frac{w_1^*}{2}$, $\frac{w_2^*}{2}$, ..., $\frac{w_{l}^*}{2}$) will be an equilibrium. To this end, we have to solve the following task:
\begin{equation}
P(\textbf{s}^*, \textbf{t}^*) = \underset{\textbf{w}}{\text{max}}~\sum_{i=1}^{l}n\frac{w_i}{2}\left(1 - H(p_i)\right)
\end{equation}
subjected to $\textbf{w}\in [0, 1]^l$ and
\begin{equation}
\sum_{i=1}^{l}n\frac{w_i}{4}\rho(\textbf{x}_i) \leq d.
\end{equation}

This is a classical \emph{linear programming} problem, which can be solved by \emph{simplex algorithm}. Accordingly, an equilibrium $\textbf{w}^*$ can be determined as a vertex of the corresponding high-dimensional polytope. Actually, when Alice and Bob share the secret key of choosing the bit-positions to be embedded, they can be considered as a whole. That means, for a strategy profile $(\textbf{s}, \textbf{t})$, essentially, it requires us to randomly choose $(s_i+t_i)\times 100\%$ of the bit-positions of $\textbf{x}_i$ for data embedding for all $1\leq i\leq l$. Thus, we only need consider one encoder. In this way, we can rewrite the payoff as:
\begin{equation}
P(\textbf{w}) = \sum_{i=1}^{l}nw_i\left(1 - H(p_i)\right),
\end{equation}
where $\textbf{w}\in [0, 1]^l$ indicates the strategy. It is required that:
\begin{equation}
\sum_{i=1}^{l}n\frac{w_i}{2}\rho(\textbf{x}_i) \leq d.
\end{equation}

Obviously, maximizing $P(\textbf{w})$ subjected to Eq. (37) is equivalent to solving Eq. (34). It is true for Case 2 that,
\begin{equation}
P(\textbf{s}^*, \textbf{t}^*) \leq \sum_{i=1}^{l}n\left(1 - H(p_i)\right),
\end{equation}
which allows that
\begin{equation}
d\leq \sum_{i=1}^{l}\frac{n}{2}\rho(\textbf{x}_i).
\end{equation}

Neither Alice or Bob deviates from an equilibrium. It is inferred that, the equilibrium $(\textbf{s}^*, \textbf{t}^*)$ should be corresponding to such a pair $(\alpha^*,\beta^*) \in [0,1]^2$ that, $s_i^* = \frac{H^{-1}\left(1-\alpha^*\rho(\textbf{x}_i)/\rho(\textbf{x}_1)\right)-p_i}{1/2-p_i}$ and $t_i^* = 2H^{-1}\left(1-H(p_i)-\beta^*\rho(\textbf{x}_i)/\rho(\textbf{x}_1)\right)$, where
\begin{equation}
\sum_{i=1}^{l}\frac{ns_i^*}{2}\rho(\textbf{x}_i) \leq d~\text{and}~\sum_{i=1}^{l}\frac{nt_i^*}{2}\rho(\textbf{x}_i)\leq d.
\end{equation}

This can be effectively solved by \emph{triple search}\footnote{Triple search allows us to find the minimum or maximum point of a parabola within a time complexity of $O(\text{log}_3N)$.} or \emph{gradient descent} since Eq. (29, 30, 31) show a \emph{concave payoff function}. Notice that, here, we do not require that both Alice and Bob introduce the maximum distortion $d$.

\section{Conclusion}
We present game-theoretic analysis to content-adaptive reversible watermarking. Different from games designed for steganography and robust watermarking, which aim to find the equilibrium between the encoder and the attacker, we focus on two encoders, who, however, do not intentionally attack each other, but rather compete with each other for maximizing their own embedding capacity. We show that, when both players do not cooperate with each other, the optimal response for each player, surprisingly, depends on his or her own distortion constraint. When they cooperate with each other, the equilibrium depends on the side information shared between them. We have analyzed two different cases for the cooperative game and provided the effective ways to find the equilibriums.

\end{document}